\documentclass[reprint,amsmath,amssymb,aps,prd]{revtex4-2}

\usepackage[utf8]{inputenc}
\usepackage{hyperref}
\usepackage{microtype}
\usepackage{scalerel}
\usepackage{tikz}
\usetikzlibrary{svg.path}
\usepackage[normalem]{ulem}
\usepackage{xcolor}

\definecolor{orcidlogocol}{HTML}{A6CE39}
\tikzset{
  orcidlogo/.pic={
    \fill[orcidlogocol] svg{M256,128c0,70.7-57.3,128-128,128C57.3,256,0,198.7,0,128C0,57.3,57.3,0,128,0C198.7,0,256,57.3,256,128z};
    \fill[white] svg{M86.3,186.2H70.9V79.1h15.4v48.4V186.2z}
                 svg{M108.9,79.1h41.6c39.6,0,57,28.3,57,53.6c0,27.5-21.5,53.6-56.8,53.6h-41.8V79.1z M124.3,172.4h24.5c34.9,0,42.9-26.5,42.9-39.7c0-21.5-13.7-39.7-43.7-39.7h-23.7V172.4z}
                 svg{M88.7,56.8c0,5.5-4.5,10.1-10.1,10.1c-5.6,0-10.1-4.6-10.1-10.1c0-5.6,4.5-10.1,10.1-10.1C84.2,46.7,88.7,51.3,88.7,56.8z};
  }
}

\newcommand\orcidicon[1]{\href{https://orcid.org/#1}{\mbox{\scalerel*{
\begin{tikzpicture}[yscale=-1,transform shape]
\pic{orcidlogo};
\end{tikzpicture}
}{|}}}}

\begin{document}

\title{Curvature-induced phantom mimicry and cosmological bounces without violating the Dominant Energy Condition}

\author{Miguel Cruz\orcidicon{0000-0003-3826-1321}}
\email{miguelcruz02@uv.mx}
\affiliation{Facultad de Física, Universidad Veracruzana, 91097 Xalapa, Veracruz, México}

\author{Samuel Lepe\orcidicon{0000-0002-3464-8337}}
\email{samuel.lepe@pucv.cl}
\affiliation{Instituto de Física, Pontificia Universidad Católica de Valparaíso, Casilla 4950, Valparaíso, Chile}

\begin{abstract}
We explore the active geometric contribution of spatial curvature in a closed universe, distinguishing between kinematic superacceleration and the strict violation of matter energy conditions. We demonstrate that for pressureless matter obeying the Dominant Energy Condition (DEC), positive spatial curvature does not induce a phantom regime, but leads to a decelerating expansion that culminates in a late-time turnaround ($\dot{H}<0$). At this finite-scale moment, the expansion ceases and the apparent horizon stays regular, signaling the beginning of an eventual Big Crunch. We contrast this behavior with the genuine phantom case ($\omega<-1$), where positive curvature introduces a distinct, early finite-radius expansion bounce ($\dot{H}>0$) that cannot prevent the inevitable late-time Big Rip singularity. Finally, we contextualize these results, discussing how positive spatial curvature can dynamically mimic phantom phenomenology and permit geodesically complete cosmologies without relying on exotic, energy-condition-violating matter.
\end{abstract}

\maketitle

\section{Introduction}
In standard relativistic cosmology, the conventional approach to understanding gravitational attraction and the causal limits of matter relies heavily on the classical energy conditions \cite{Kontou2020}. For geodesic motion, the Raychaudhuri equation implies that two neighboring observers will converge if $V_{a}R^{a}{}_{b}V^{b} \geq 0$. This inequality preserves the attractive character of gravity and is widely known as the timelike convergence condition, where $V^{a}$ is the unit vector tangent to the congruence and $R_{ab}$ is the Ricci tensor. Crucially, these constraints are always violated during any epoch of cosmic accelerated expansion \cite{Riess1998,Perlmutter1999}.

For a standard phantom fluid, defined by an equation of state of the form $\omega \equiv p/\rho < -1$ \cite{Caldwell2002}, these foundational restrictions are completely broken, inevitably leading to a catastrophic Big Rip singularity where the scale factor, energy density, and Hubble parameter diverge simultaneously in finite proper time \cite{Caldwell2003,Nojiri2005}.

While energy conditions strictly constrain the physical matter sector ($T_{ab}$), the ultimate destiny of cosmic evolution is governed by the full geometric structure of the gravitational field equations. Within an extended framework, the field equations can be formally written as
\begin{equation}
G_{\mu \nu} = T^{(\mathrm{m})}_{\mu \nu}+T^{(\mathrm{eff})}_{\mu \nu},\label{eq:beyond}
\end{equation}
where $T^{(\mathrm{m})}_{\mu \nu}$ is the energy-momentum tensor associated with standard matter fields, and $T^{(\mathrm{eff})}_{\mu \nu}$ encapsulates the effective geometric contributions. This formulation naturally motivates viewing spatial curvature not simply as a passive topological parameter, but as an active geometric contribution to $T^{(\mathrm{eff})}_{\mu \nu}$.

It has been established that positive spatial curvature can support cosmological bounces that preserve the necessary energy conditions \cite{MolinaParis1998}. More recently, Burwig and Easson have extensively demonstrated that positive curvature allows for non-singular, geodesically complete cosmologies that naturally mimic phantom dark energy phenomenology without invoking exotic matter \cite{BurwigEasson2025,BurwigEasson2026,Easson2026}. In this Letter, we complement these findings by explicitly contrasting the geometry-induced turnaround of DEC-preserving fluids with the early bounce and late Big Rip of genuine phantom fluids, providing a detailed analysis of the kinematics and the role of the apparent horizon at these critical points. Throughout, we adopt units $8\pi G=c=1$ and normalize the scale factor as $x\equiv a/a_{0}$.

\section{The Deceleration Parameter and Energy Conditions}
To distinguish between necessary and sufficient constraints, it is helpful to briefly formalize the classical energy conditions for a perfect fluid with density $\rho$ and pressure $p = \omega\rho$:
\begin{itemize}
    \item \textbf{Strong Energy Condition (SEC):} Requires $\rho+p \ge 0$ and $\rho+3p \ge 0$ (i.e., $\omega \ge -1/3$).
    \item \textbf{Weak Energy Condition (WEC):} Ensures energy density measured by any observer is non-negative, requiring $\rho \ge 0$ and $\rho+p \ge 0$ ($\omega \ge -1$).
    \item \textbf{Null Energy Condition (NEC):} The limit of the WEC for null observers, demanding $\rho+p \ge 0$ ($\omega \ge -1$).
    \item \textbf{Dominant Energy Condition (DEC):} Establishes that energy cannot flow faster than light, requiring $\rho \ge 0$ and the strict bound $|p| \le \rho$. Thus, the full DEC dictates $-1 \le \omega \le 1$.
\end{itemize}

By considering the standard Friedmann constraint and the fluid conservation equation in the presence of curvature,
\begin{equation}
3H^{2}=\rho-\frac{3k}{a^{2}}, \quad \dot{\rho}+3H(1+\omega)\rho=0,\label{eq:friedmann}
\end{equation}
the matter parameter state $\omega$ can be written as a function of the kinematic deceleration parameter $q \equiv -a\ddot{a}/\dot{a}^2$ and the dimensionless curvature density parameter $\Omega_k \equiv -k/(a^2 H^2)$
\begin{equation}
\omega=\frac{2}{3}\left(\frac{q}{1-\Omega_{k}}-\frac{1}{2}\right). \label{eq:omega_q}
\end{equation}
Inverting this expression yields the deceleration parameter explicitly in terms of the physical fluid and geometry as follows, $q=(1/2)(1+3\omega)(1-\Omega_{k})$. If we enforce the NEC ($\omega \geq -1$) as a necessary condition for standard matter, Eq.~(\ref{eq:omega_q}) yields a geometric kinematic constraint previously identified by Lima, Vitenti, and Rebouças \cite{Lima2008}:
\begin{equation}
1+q \geq \Omega_{k} \implies 1+q \geq -\frac{k}{a^{2}H^{2}}.
\end{equation}
For a closed universe geometry characterized by a positive spatial curvature ($k=+1$), this inequality simplifies to:
\begin{equation}
q \geq -\left(1+\frac{1}{a^{2}H^{2}}\right).
\end{equation}
This relation formally permits the kinematic deceleration parameter to fall below $-1$. However, describing this strictly as a phantom regime requires careful evaluation of explicit solutions, as the physical nature of these critical points depends entirely on the dynamics of $\dot{H}$.



\section{Turnarounds and Bounces}

\subsection{Pressureless Matter Turnaround}
It is well known that kinematic superacceleration ($q < -1$) can occur without violating the DEC. A classic example is the expanding branch of the closed slicing of de Sitter space ($\omega=-1$), which yields $q=-\coth^{2}(H_{\Lambda}t) < -1$ for $t>0$ following a bounce at $t=0$.

To evaluate the consequences of this geometric mechanism, we examine a closed universe ($k=+1$) dominated by standard pressureless matter ($\omega=0$), which trivially satisfies the full DEC. Introducing the normalized scale factor $x \equiv a/a_0$, the corresponding Hubble evolution is described by:
\begin{equation}
H(x)=\sqrt{\frac{\rho_{0}}{3}}\frac{1}{x}\sqrt{x^{-1}-x_{s}^{-1}}, \label{eq:H_dust}
\end{equation}
where we have defined the specific turning point $x_s \equiv \theta_k^{-1}$ via the curvature parameter $\theta_k \equiv 3k/(\rho_0 a_0^2)$. To maintain a physically meaningful, real-valued Hubble parameter ($H(x) \ge 0$), the evolution of the scale factor is strictly bounded to the domain $x \le x_s$. As a result, the universe encounters a late-time event at $x = x_s$, at which the Hubble parameter vanishes $H(x_s) = 0$. 

The internal dynamics approaching this state are revealed by the exact forms of the deceleration parameter $q(x)$ and the cosmic acceleration $\dot{H}(x)$. Using $q(x) = x_s / 2(x_s - x)$, it is clear that for $x \le x_s$, the deceleration parameter is strictly positive ($q(x) > 0$) throughout the expanding branch. Furthermore, the time derivative of the Hubble parameter is given by:
\begin{equation}
\dot{H}(x) = -\frac{\rho_{0}}{3}\frac{\left(\frac{3}{2}x_{s} - x\right)}{x^{3}x_{s}} < 0.
\end{equation}
At the critical point $H=0$, applying the Friedmann constraint yields a strictly negative value, as follows
\begin{equation}
\dot{H}(x_s) = -\frac{1}{2a_s^2} < 0.
\end{equation}
Therefore, the divergence of $q(x \to x_s) \to +\infty$ simply reflects the division by $H^2=0$, not a spacetime singularity. The solution is always decelerating and never enters a phantom-like regime. Instead, it defines a smooth expansion-to-contraction turnaround, marking the maximum scale factor before an eventual Big Crunch.

\subsection{Genuine Phantom Dynamics and the Big Rip}
It is illuminating to contrast this geometric mechanism with the case of a \emph{genuine} phantom fluid ($\omega<-1$) evolving on the same closed background. Writing $\rho=\rho_{0}\,x^{3(|\omega|-1)}$ and inserting it into the Friedmann constraint yields
\begin{equation}
H(x)=\sqrt{\frac{\rho_{0}}{3}}\,\frac{1}{x}\sqrt{x^{\eta}-x_{s}^{\eta}},\qquad \eta\equiv 3|\omega|-1,\label{eq:H_phantom}
\end{equation}
with the turning point $x_{s}=\theta_{k}^{1/\eta}$. For this case, the reality condition of $H$ demands $x \ge x_{s}$. The singularity $H(x_{s})=0$ thus becomes an \emph{early} event. Evaluating the derivative of the Hubble parameter yields:
\begin{equation}
\dot{H}(x) = \frac{\rho_0}{2}\left[(|\omega|-1)(x^{\eta}-x_s^{\eta}) + \left(|\omega|-\frac{1}{3}\right)x^{\eta}\right]\frac{1}{x^2}.
\end{equation}
At the critical scale factor $x_s$, this reduces to:
\begin{equation}
\dot{H}(x_s) = \frac{\rho_0}{2}\left(|\omega|-\frac{1}{3}\right)x_s^{\eta-2} > 0.
\end{equation}
The strictly positive derivative identifies a true cosmological bounce from which expansion proceeds, accompanied by a deceleration parameter that diverges negatively, $q(x_s) \to -\infty$.

However, positive curvature introduces this finite-radius bounce separately from the ultimate cosmic fate. As the scale factor grows ($x \gg x_s$), the curvature term in the Friedmann equation becomes completely negligible, and $H^2(x) \simeq (\rho_0/3)x^{3(|\omega|-1)}$. In this limit:
\begin{equation}
H(x \to \infty) \to \infty, \quad \rho(x \to \infty) \to \infty, \quad q \to -(3|\omega|-1).
\end{equation}
Thus, a genuine phantom fluid on a closed background interpolates between an early bounce and a late Big Rip. The curvature does not displace or prevent the Big Rip at infinite scale factors, as simultaneous divergences are the defining behavior of this singularity.

\section{Apparent Horizon and the Physical Interpretation}
A sharper physical reading of the event $H(x_{s})=0$ follows from the behavior of the apparent horizon. In the presence of curvature the relevant apparent horizon length scale is
\begin{equation}
r_{\mathrm{app}}=\frac{1}{\sqrt{H^{2}+k/a^{2}}},\label{eq:rH}
\end{equation}
which reduces to the standard Hubble radius $1/|H|$ only for $k=0$. Using the Friedmann equation, the apparent-horizon radius is strictly defined by the energy density: $r_{\mathrm{app}} = \sqrt{3/\rho}$.

When the expansion ceases and $H \to 0$, the usual Hubble radius $1/|H|$ becomes infinite. Nevertheless, Eq. (\ref{eq:rH}) yields
\begin{equation}
r_{\mathrm{app}}(a_{s})=\frac{a_{s}}{\sqrt{k}}.
\end{equation}
For a closed geometry ($k=+1$), the apparent horizon stays finite and non-zero, directly reflecting the finite, positive matter density at this epoch. The regularity of the apparent horizon confirms that the point $H(x_{s})=0$ represents a non-singular turning point (or bounce) rather than a disruptive curvature singularity.

\section{Discussion: Phantom Mimicry and Geodesic Completeness}
The mechanism explored here is intrinsically tied to a closed geometry. The ability of positive curvature to re-scale the standard cosmological equations provides a robust framework for phantom-mimicry. As highlighted by Easson and Burwig \cite{BurwigEasson2025}, the Friedmann equations imply that:
\begin{equation}
\dot{H} = -\frac{1}{2}(\rho+p) + \frac{k}{a^2}.
\end{equation}
In a closed universe ($k=+1$), this structure permits $\dot{H} > 0$ even when the NEC is fully satisfied ($\rho+p \ge 0$). Thus, positive spatial curvature can produce apparently phantom behavior without the need for exotic, energy-condition-violating matter.

Furthermore, this geometric mechanism is central to avoid spacetime singularities. Burwig and Easson have recently shown that while flat and open FLRW cosmologies are obstructed by the affine-Averaged Null Energy Condition (ANEC), closed FLRW branches uniquely escape these restrictions, allowing for geodesically complete bouncing cosmologies \cite{BurwigEasson2026,Easson2026}. Although our explicit pressureless matter solution yields a decelerating turnaround rather than a superaccelerating bounce, it underscores that structural geometric components, rather than just material content, dictate the ultimate presence of singularities and turnarounds.

\section{Conclusion}
The standard paradigm in modern cosmology widely treats phantom expansion and future singularities as synonymous with the violation of classical energy conditions. In this Letter, we have explored the active geometric contribution of positive spatial curvature ($k=+1$). By rigorously distinguishing between kinematic and fluid restrictions, we demonstrated that an explicit closed universe with pressureless matter strictly obeying the Dominant Energy Condition does not enter a phantom regime. Instead, it decelerates into a late-time turnaround ($\dot{H}<0$) free of curvature singularities, paving the way for contraction.

We contrasted this outcome with a genuine phantom fluid ($\omega<-1$), where positive curvature introduces a distinct, early finite-radius bounce ($\dot{H}>0$), but ultimately cannot prevent the asymptotic infinite-scale Big Rip. At these finite turning points ($H=0$), it is the apparent horizon-not the Hubble radius-that remains finite, reflecting a bounded and regular energy density. As demonstrated in recent literature, the intrinsic capacity of positive spatial curvature to support $\dot{H}>0$ dynamics without exotic matter provides a natural mechanism for phantom mimicry, establishing closed geometries as a critical avenue for exploring geodesically complete, non-singular cosmologies.

\section*{Acknowledgments}
M.~Cruz work was partially supported by S.N.I.I. (SECIHTI-M\'exico). S.~Lepe acknowledges the FONDECYT grant N°1250969, Chile.


\appendix
\section{Cosmic time of the bounces}
\label{sec:app}
The cosmic time up to a scale factor $x$ follows from $t=\int dx/(xH)$, with $xH=\sqrt{\rho_{0}/3}\,\sqrt{x^{-1}-\theta_{k}}$ for pressureless matter and $xH=\sqrt{\rho_{0}/3}\,\sqrt{x^{\eta}-\theta_{k}}$ for the phantom fluid Eq.~(\ref{eq:H_phantom}). Both integrals share the structure $_{2}F_{1}(\tfrac{1}{2},b;b+1;z)$; introducing the shorthand notation
\begin{equation}
\mathcal{F}(b;z)\equiv{}_{2}F_{1}\!\left(\tfrac{1}{2},b;b+1;z\right),
\end{equation}
the late turnaround ($\omega=0$, $x_{s}=\theta_{k}^{-1}$) is described by
\begin{align}
\sqrt{\tfrac{\rho_{0}}{3}}\,t&=\frac{2}{3}\,x^{3/2}\,\mathcal{F}\!\left(\tfrac{3}{2};\theta_{k}x\right)+C,\\
t_{s}&=t_{0}+\frac{2}{3}\sqrt{\tfrac{3}{\rho_{0}}}\left[x_{s}^{3/2}\,\mathcal{F}\!\left(\tfrac{3}{2};1\right)-\mathcal{F}\!\left(\tfrac{3}{2};x_{s}^{-1}\right)\right]>t_{0},
\end{align}
and the early bounce ($\omega<-1$, $x_{s}=\theta_{k}^{1/\eta}$, $\eta=3|\omega|-1$, $b_{\eta}\equiv\tfrac{\eta-2}{2\eta}$) by
\begin{align}
\sqrt{\tfrac{\rho_{0}}{3}}\,t&=\frac{2}{2-\eta}\,x^{(2-\eta)/2}\,\mathcal{F}\!\left(b_{\eta};\theta_{k}x^{-\eta}\right)+C,\\
t_{s}&=t_{0}+\frac{2}{2-\eta}\sqrt{\tfrac{3}{\rho_{0}}}\left[x_{s}^{(2-\eta)/2}\,\mathcal{F}\!\left(b_{\eta};1\right)-\mathcal{F}\!\left(b_{\eta};\theta_{k}\right)\right]<t_{0}.
\end{align}
The first parameter of $_{2}F_{1}$ is $+1/2$, required to reproduce the integrand; since $c-a-b=\tfrac{1}{2}>0$ in both cases, $\mathcal{F}(b;1)$ converges and each $t_{s}$ is finite-occurring after ($\omega=0$) or before ($\omega<-1$) the present epoch.

\end{document}